\documentclass[conference]{IEEEtran}
\usepackage{cite}

\ifCLASSINFOpdf
\usepackage[pdftex]{graphicx}
\else
\fi
\usepackage{amsmath}
\usepackage{amsfonts} 
\usepackage{amsthm}
\usepackage{amssymb}
\usepackage[boxruled]{algorithm2e}
\usepackage{amsthm} 
\usepackage{mathtools} 

\usepackage{BOONDOX-ds}

\usepackage{bbm}

\newcommand{\pfun}{\mathop{\hbox{$\to$\kern-7pt\raise.9pt\hbox{\scalebox{1}[.55]{$|$}}\kern4pt} }}

\usepackage{array}

\begin{document}

\title{A Billion Updates per Second Using 30,000 Hierarchical In-Memory D4M Databases}

\author{\IEEEauthorblockN{Jeremy Kepner$^{1,2,3}$, Vijay Gadepally$^{1,2}$, Lauren
    Milechin$^4$, Siddharth Samsi$^1$, \\ William Arcand$^1$, David  Bestor$^1$, William Bergeron$^1$, 
Chansup Byun$^1$, Matthew Hubbell$^1$, \\ Micheal Houle$^1$, Micheal Jones$^1$, Anne Klein$^1$, Peter Michaleas$^1$, \\ Julie Mullen$^1$, Andrew Prout$^1$, Antonio Rosa$^1$,  Charles
Yee$^1$, Albert Reuther$^1$
\\
\IEEEauthorblockA{$^1$MIT Lincoln Laboratory Supercomputing Center, $^2$MIT Computer Science \& AI Laboratory, \\ $^3$MIT Mathematics Department, $^4$MIT Department of Earth, Atmospheric and Planetary Sciences}}}
\maketitle

\begin{abstract}
Analyzing large scale networks requires high performance streaming updates of graph representations of these data.  Associative arrays are mathematical objects combining properties of spreadsheets, databases, matrices, and graphs, and are well-suited for representing and analyzing streaming network data.  The Dynamic Distributed Dimensional Data Model (D4M) library implements associative arrays in a variety of languages (Python, Julia, and Matlab/Octave) and provides a lightweight in-memory database.  Associative arrays are designed for block updates. Streaming updates to a large associative array requires a hierarchical implementation to optimize the performance of the memory hierarchy.  Running 34,000 instances of a hierarchical D4M associative arrays on 1,100 server nodes on the MIT SuperCloud achieved a sustained update rate of 1,900,000,000 updates per second.  This capability allows the MIT SuperCloud to analyze extremely large streaming network data sets. 
\end{abstract}

%
\IEEEpeerreviewmaketitle

\section{Introduction}
\let\thefootnote\relax\footnotetext{This material is based upon work supported by the Assistant Secretary of Defense for Research and Engineering under Air Force Contract No. FA8702-15-D-0001 and National Science Foundation grants DMS-1312831 and CCF-1533644. Any opinions, findings, conclusions or recommendations expressed in this material are those of the author(s) and do not necessarily reflect the views of the Assistant Secretary of Defense for Research and Engineering or the National Science Foundation.}

\begin{figure*}[t]
\centering
\includegraphics[width=6in]{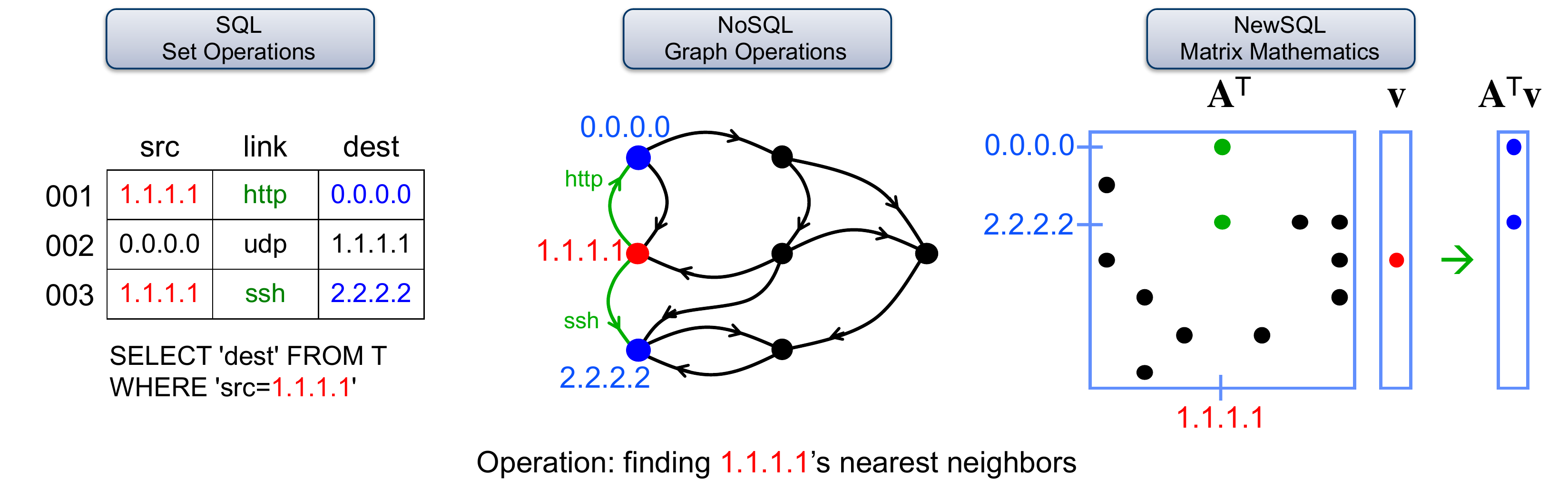}
\caption{Associative arrays combine the properties of databases, graphs, and matrices and provide common mathematics that span SQL, NoSQL, and NewSQL databases, and are ideal for analyzing networks.  The diagram shows the graph operation of finding the neighbors of 1.1.1.1 in each representation.}
\label{fig:AssociativeArrays}
\end{figure*}

\begin{figure*}[]
\centering
\includegraphics[width=6in]{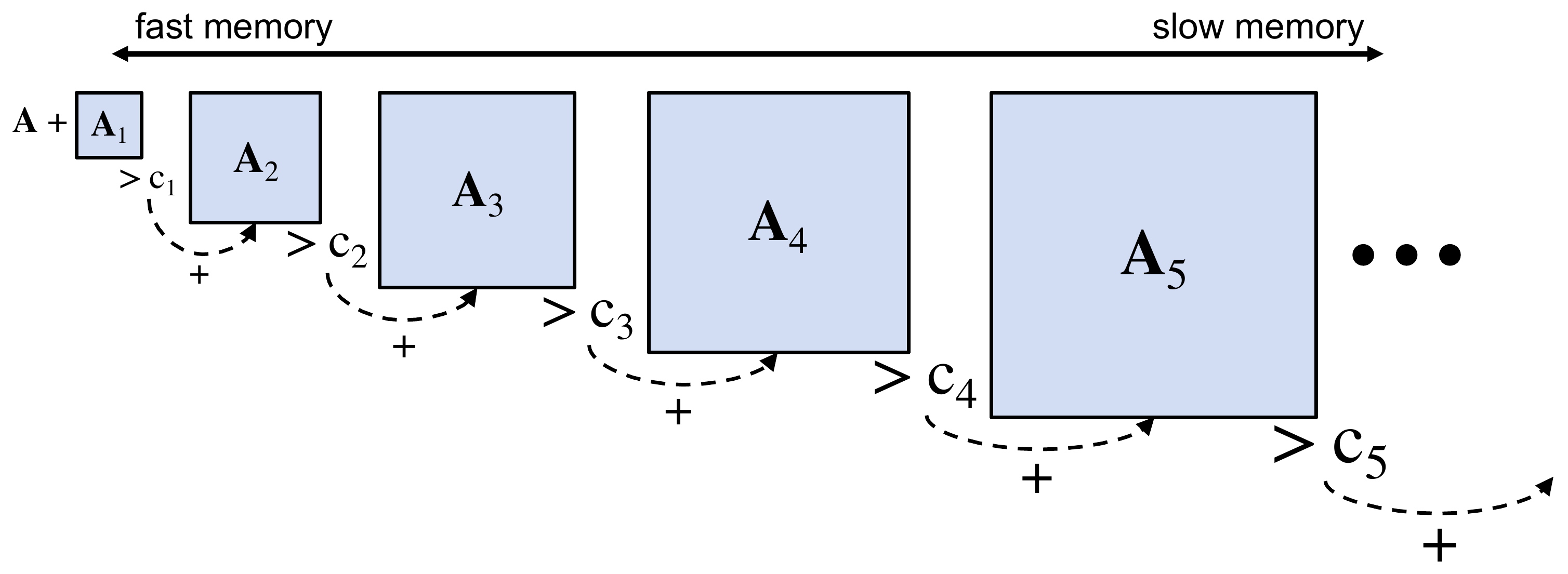}
\caption{Hierarchical associative arrays store increasing numbers of non-zero entries in each layer.  If layer ${\bf A}_i$ surpasses the non-zero threshold $c_i$ it is added to ${\bf A}_{i+1}$ and cleared.  Hierarchical arrays ensure that the majority of updates are performed in fast memory.}
\label{fig:HierarchicalArrays}
\end{figure*}

Networks form the basis of worldwide communication and it is estimated that in 2018, there will be almost 37 Terabytes per second  (TB/s) of Internet Protocol (IP) traffic.
The rapid rise of sophisticated cyber threats is well documented and a growing threat to our information systems~\cite{kshetri2009positive,hale2002cybercrime}.
Development of novel computer network traffic analytics requires: high level programming environments, massive amount of network data, and diverse data products for ``at scale'' algorithm pipeline development.  Our team has developed a scalable network analytics platform applied to a network data using the D4M (Dynamic Distributed Dimensional Data Model) analytics environment and MIT SuperCloud interactive computing environment \cite{gadepally2018hyperscaling}. D4M combines the power of sparse linear algebra, associative arrays, parallel processing, and distributed databases (such as SciDB and Apache Accumulo) to provide a scalable data and computation system that addresses the big data problems associated with network analytics development. The MIT SuperCloud allows users to interactively process massive amounts of data in minutes on many thousands of cores using the software and environments most familiar to them.  A key challenge for this pipeline is handling streaming updates of a large networks.  This paper describes the implementation of a hierarchical approach designed to optimize the performance of the memory hierarchy.

\section{Hierarchical Associative Arrays}

Analyzing large scale networks requires high performance streaming updates of graph representations of these data.  Associative arrays are mathematical objects combining properties of spreadsheets, databases, matrices, and graphs, and are well-suited for representing and analyzing streaming network data (see Fig.~\ref{fig:AssociativeArrays}). 
In many databases, these table operations can be mapped onto well-defined mathematical operations with known mathematical properties.  For example, relational (or SQL) databases \cite{Stonebraker1976,date1989guide,elmasri2010fundamentals} are described by relational algebra \cite{codd1970relational,maier1983theory,Abiteboul1995} that corresponds to the union-intersection semiring ${\cup}.{\cap}$ \cite{jananthan2017polystore}.  Triple-store databases (NoSQL) \cite{DeCandia2007,LakshmanMalik2010,George2011,Wall2015} and analytic databases (NewSQL) \cite{Stonebraker2005,Kallman2008,Balazinska2009,StonebrakerWeisberg2013,Hutchison2015,gadepally2015graphulo}  follow similar mathematics \cite{kepner2016associative}.  The table operations of these databases are further encompassed by associative array algebra, which brings the beneficial properties of matrix mathematics and sparse linear systems theory, such as closure, commutativity, associativity, and distributivity \cite{kepnerjananthan}.
The aforementioned mathematical properties provide strong correctness guarantees that are independent of scale and particularly helpful when trying to reason about massively parallel systems.


The D4M library implements associative arrays in a variety of languages (Python, Julia, and Matlab/Octave) and provides a lightweight in-memory database.  Associative arrays are designed for block updates. Streaming updates to a large associative array requires a hierarchical implementation to optimize the performance of the memory hierarchy (see Fig.~\ref{fig:HierarchicalArrays}).   Rapid updates are performed on the smallest arrays in the fastest memory.  If the number of entries exceeds the threshold $c_i$, then ${\bf A}_i$ is added to ${\bf A}_{i+1}$ and ${\bf A}_i$ is cleared.  Hierarchical arrays dramatically reduce the number of updates to slow memory.  Upon query, all layers in the hierarchy are summed into largest array.  The cut values $c_i$ can be selected so as to optimize the performance with respect to particular applications.

\section{Performance Results}

\begin{figure}[]
\centering
\includegraphics[width=\columnwidth]{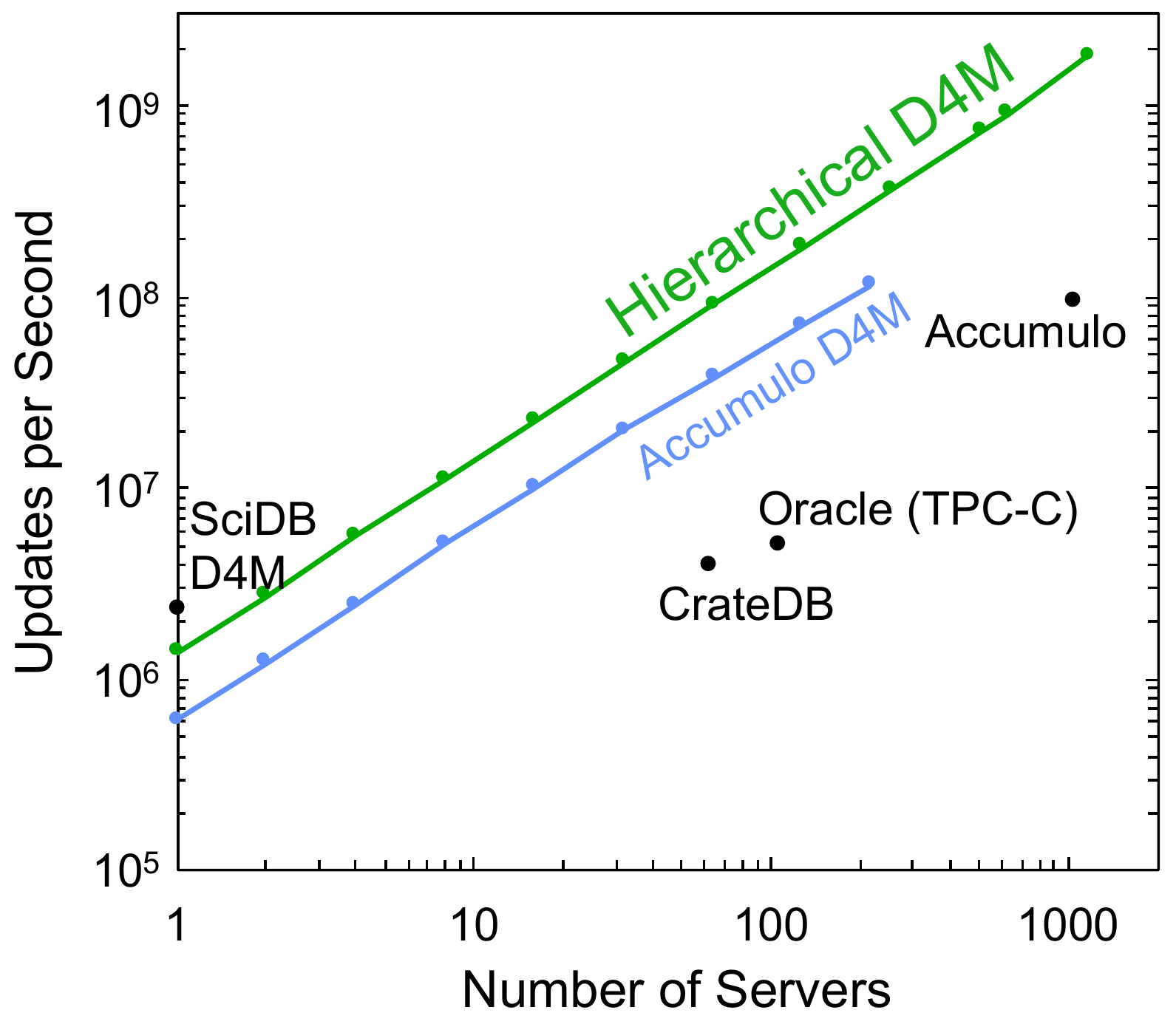}
\caption{Update rate as a function of number of servers for hierarchical D4M associative arrays and other previous published work: Accumulo D4M \cite{kepner2014achieving}, SciDB D4M \cite{samsi2016benchmarking}, Accumulo \cite{sen2013benchmarking}, Oracle TPC-C benchmark, and CrateDB \cite{CrateDB}}
\label{fig:UpdateRate}
\end{figure}

  The performance of associative arrays are benchmarked using a power-law graph of 100,000,000 entries divided up into 1,000 sets of 100,000 entries.  These data were then simultaneously loaded and updated using varying numbers processes on varying number of nodes on the MIT SuperCloud. This experiment mimics thousands of processors each creating many different graphs of 100,000,000 edges each.  In a real analysis application, each process would also compute various network statistics on each of the streams as they are updated.   The update rate as function of severs nodes is shown on Fig.~\ref{fig:UpdateRate}.  The achieved update rate of 1,900,000,000 updates per second is significantly larger than prior published results. This capability allows the MIT SuperCloud to analyze extremely large streaming network data sets.


%
%




\bibliographystyle{ieeetr}

\bibliography{aarabib}
%

\end{document}